\documentclass[
prl,
reprint,
superscriptaddress, 
 amsmath,amssymb,
 aps,
]{revtex4-2}
\usepackage{graphicx} 
\usepackage{amsmath}
\usepackage{xspace}
\usepackage{dcolumn}
\usepackage{bm}

\newcommand{\SK}{Super-K\xspace}
\newcommand{\smhigh}{B16+GS98\xspace}
\newcommand{\smlow}{B16+AGSS09\xspace}

\newcommand{\czvar}{K2-A09-12\xspace}
\newcommand{\mzvar}{K2-A09-MZvar-12\xspace}

\newcommand{\sno}{SNO\xspace}
\newcommand{\pp}{pp chain\xspace}

\begin{document}

\preprint{APS/123-QED}

\title{$^8$B Solar Neutrino Measurements of Solar Core Size and Temperature\\ from 5805 Days of Super-Kamiokande}

\newcommand{\AFFicrr}{\affiliation{Kamioka Observatory, Institute for Cosmic Ray Research, University of Tokyo, Kamioka, Gifu 506-1205, Japan}}
\newcommand{\AFFkashiwa}{\affiliation{Research Center for Cosmic Neutrinos, Institute for Cosmic Ray Research, University of Tokyo, Kashiwa, Chiba 277-8582, Japan}}
\newcommand{\AFFipmu}{\affiliation{Kavli Institute for the Physics and
Mathematics of the Universe (WPI), The University of Tokyo Institutes for Advanced Study,
University of Tokyo, Kashiwa, Chiba 277-8583, Japan }}
\newcommand{\AFFmad}{\affiliation{Department of Theoretical Physics, University Autonoma Madrid, 28049 Madrid, Spain}}
\newcommand{\AFFubc}{\affiliation{Department of Physics and Astronomy, University of British Columbia, Vancouver, BC, V6T1Z4, Canada}}
\newcommand{\AFFbu}{\affiliation{Department of Physics, Boston University, Boston, MA 02215, USA}}
\newcommand{\AFFuci}{\affiliation{Department of Physics and Astronomy, University of California, Irvine, Irvine, CA 92697-4575, USA }}
\newcommand{\AFFcsu}{\affiliation{Department of Physics, California State University, Dominguez Hills, Carson, CA 90747, USA}}
\newcommand{\AFFcnm}{\affiliation{Institute for Universe and Elementary Particles, Chonnam National University, Gwangju 61186, Korea}}
\newcommand{\AFFduke}{\affiliation{Department of Physics, Duke University, Durham NC 27708, USA}}
\newcommand{\AFFgifu}{\affiliation{Department of Physics, Gifu University, Gifu, Gifu 501-1193, Japan}}
\newcommand{\AFFgist}{\affiliation{GIST College, Gwangju Institute of Science and Technology, Gwangju 500-712, Korea}}
\newcommand{\AFFuh}{\affiliation{Department of Physics and Astronomy, University of Hawaii, Honolulu, HI 96822, USA}}
\newcommand{\AFFicl}{\affiliation{Department of Physics, Imperial College London , London, SW7 2AZ, United Kingdom }}
\newcommand{\AFFkek}{\affiliation{High Energy Accelerator Research Organization (KEK), Tsukuba, Ibaraki 305-0801, Japan }}
\newcommand{\AFFkobe}{\affiliation{Department of Physics, Kobe University, Kobe, Hyogo 657-8501, Japan}}
\newcommand{\AFFkyoto}{\affiliation{Department of Physics, Kyoto University, Kyoto, Kyoto 606-8502, Japan}}
\newcommand{\AFFliv}{\affiliation{Department of Physics, University of Liverpool, Liverpool, L69 7ZE, United Kingdom}}
\newcommand{\AFFmiyagi}{\affiliation{Department of Physics, Miyagi University of Education, Sendai, Miyagi 980-0845, Japan}}
\newcommand{\AFFnagoya}{\affiliation{Institute for Space-Earth Environmental Research, Nagoya University, Nagoya, Aichi 464-8602, Japan}}
\newcommand{\AFFkmi}{\affiliation{Kobayashi-Maskawa Institute for the Origin of Particles and the Universe, Nagoya University, Nagoya, Aichi 464-8602, Japan}}
\newcommand{\AFFpol}{\affiliation{National Centre For Nuclear Research, 02-093 Warsaw, Poland}}
\newcommand{\AFFsuny}{\affiliation{Department of Physics and Astronomy, State University of New York at Stony Brook, NY 11794-3800, USA}}
\newcommand{\AFFokayama}{\affiliation{Department of Physics, Okayama University, Okayama, Okayama 700-8530, Japan }}
\newcommand{\AFFosaka}{\affiliation{Department of Physics, Osaka University, Toyonaka, Osaka 560-0043, Japan}}
\newcommand{\AFFox}{\affiliation{Department of Physics, Oxford University, Oxford, OX1 3PU, United Kingdom}}
\newcommand{\AFFqmul}{\affiliation{School of Physics and Astronomy, Queen Mary University of London, London, E1 4NS, United Kingdom}}
\newcommand{\AFFregina}{\affiliation{Department of Physics, University of Regina, 3737 Wascana Parkway, Regina, SK, S4SOA2, Canada}}
\newcommand{\AFFseoul}{\affiliation{Department of Physics and Astronomy, Seoul National University, Seoul 151-742, Korea}}
\newcommand{\AFFsheff}{\affiliation{School of Mathematical and Physical Sciences, University of Sheffield, S3 7RH, Sheffield, United Kingdom}}
\newcommand{\AFFshizuokasc}{\affiliation{Department of Informatics in
Social Welfare, Shizuoka University of Welfare, Yaizu, Shizuoka, 425-8611, Japan}}
\newcommand{\AFFstfc}{\affiliation{STFC, Rutherford Appleton Laboratory, Harwell Oxford, and Daresbury Laboratory, Warrington, OX11 0QX, United Kingdom}}
\newcommand{\AFFskk}{\affiliation{Department of Physics, Sungkyunkwan University, Suwon 440-746, Korea}}
\newcommand{\AFFtodai}{\affiliation{Department of Physics, University of Tokyo, Bunkyo, Tokyo 113-0033, Japan }}
\newcommand{\AFFtit}{\affiliation{Department of Physics, Institute of Science Tokyo, Meguro, Tokyo 152-8551, Japan }}
\newcommand{\AFFtus}{\affiliation{Department of Physics and Astronomy, Faculty of Science and Technology, Tokyo University of Science, Noda, Chiba 278-8510, Japan }}
\newcommand{\AFFtriumf}{\affiliation{TRIUMF, 4004 Wesbrook Mall, Vancouver, BC, V6T2A3, Canada }}
\newcommand{\AFFtokai}{\affiliation{Department of Physics, Tokai University, Hiratsuka, Kanagawa 259-1292, Japan}}
\newcommand{\AFFtsinghua}{\affiliation{Department of Engineering Physics, Tsinghua University, Beijing, 100084, China}}
\newcommand{\AFFynu}{\affiliation{Department of Physics, Yokohama National University, Yokohama, Kanagawa, 240-8501, Japan}}
\newcommand{\AFFllr}{\affiliation{Ecole Polytechnique, IN2P3-CNRS, Laboratoire Leprince-Ringuet, F-91120 Palaiseau, France }}
\newcommand{\AFFbari}{\affiliation{ Dipartimento Interuniversitario di Fisica, INFN Sezione di Bari and Universit\`a e Politecnico di Bari, I-70125, Bari, Italy}}
\newcommand{\AFFnapoli}{\affiliation{Dipartimento di Fisica, INFN Sezione di Napoli and Universit\`a di Napoli, I-80126, Napoli, Italy}}
\newcommand{\AFFroma}{\affiliation{INFN Sezione di Roma and Universit\`a di Roma ``La Sapienza'', I-00185, Roma, Italy}}
\newcommand{\AFFpadova}{\affiliation{Dipartimento di Fisica, INFN Sezione di Padova and Universit\`a di Padova, I-35131, Padova, Italy}}
\newcommand{\AFFkeio}{\affiliation{Department of Physics, Keio University, Yokohama, Kanagawa, 223-8522, Japan}}
\newcommand{\AFFwinnipeg}{\affiliation{Department of Physics, University of Winnipeg, MB R3J 3L8, Canada }}
\newcommand{\AFFkcl}{\affiliation{Department of Physics, King's College London, London, WC2R 2LS, UK }}
\newcommand{\AFFwarwick}{\affiliation{Department of Physics, University of Warwick, Coventry, CV4 7AL, UK }}
\newcommand{\AFFral}{\affiliation{Rutherford Appleton Laboratory, Harwell, Oxford, OX11 0QX, UK }}
\newcommand{\AFFwu}{\affiliation{Faculty of Physics, University of Warsaw, Warsaw, 02-093, Poland }}
\newcommand{\AFFbcit}{\affiliation{Department of Physics, British Columbia Institute of Technology, Burnaby, BC, V5G 3H2, Canada }}
\newcommand{\AFFtohoku}{\affiliation{Department of Physics, Faculty of Science, Tohoku University, Sendai, Miyagi, 980-8578, Japan }}
\newcommand{\AFFicise}{\affiliation{Institute For Interdisciplinary Research in Science and Education, ICISE, Quy Nhon, 55121, Vietnam }}
\newcommand{\AFFilance}{\affiliation{ILANCE, CNRS - University of Tokyo International Research Laboratory, Kashiwa, Chiba 277-8582, Japan}}
\newcommand{\AFFibs}{\affiliation{Center for Underground Physics, Institute for Basic Science (IBS), Daejeon, 34126, Korea}}
\newcommand{\AFFglasgow}{\affiliation{School of Physics and Astronomy, University of Glasgow, Glasgow, Scotland, G12 8QQ, United Kingdom}}
\newcommand{\AFFoecu}{\affiliation{Media Communication Center, Osaka Electro-Communication University, Neyagawa, Osaka, 572-8530, Japan}}
\newcommand{\AFFminn}{\affiliation{School of Physics and Astronomy, University of Minnesota, Minneapolis, MN  55455, USA}}
\newcommand{\AFFsilesia}{\affiliation{August Che\l{}kowski Institute of Physics, University of Silesia in Katowice, 75 Pu\l{}ku Piechoty 1, 41-500 Chorz\'{o}w, Poland}}
\newcommand{\AFFtoyama}{\affiliation{Faculty of Science, University of Toyama, Toyama City, Toyama 930-8555, Japan}}
\newcommand{\AFFbmcc}{\affiliation{Science Department, Borough of Manhattan Community College / City University of New York, New York, New York, 1007, USA.}}
\newcommand{\AFFnumazu}{\affiliation{National Institute of Technology, Numazu College, Numazu, Shizuoka  410-8501, Japan}}
\newcommand{\AFFniihama}{\affiliation{National Institute of Technology, Niihama College, Niihama, Ehime  792-8580, Japan}}
\newcommand{\AFFucas}{\affiliation{School of Physical Sciences, University of Chinese Academy of Sciences, Beijing 101408, China}}

\AFFicrr
\AFFkashiwa
\AFFmad
\AFFbmcc
\AFFbu
\AFFbcit
\AFFuci
\AFFcsu
\AFFucas
\AFFcnm
\AFFduke
\AFFllr
\AFFgifu
\AFFgist
\AFFglasgow
\AFFuh
\AFFibs
\AFFicise
\AFFicl
\AFFbari
\AFFnapoli
\AFFpadova
\AFFroma
\AFFilance
\AFFkeio
\AFFkek
\AFFkcl
\AFFkobe
\AFFkyoto
\AFFliv
\AFFminn
\AFFmiyagi
\AFFnagoya
\AFFkmi
\AFFpol
\AFFniihama
\AFFnumazu
\AFFsuny
\AFFokayama
\AFFoecu
\AFFox
\AFFral
\AFFseoul
\AFFsheff
\AFFshizuokasc
\AFFsilesia
\AFFstfc
\AFFskk
\AFFtohoku
\AFFtodai
\AFFipmu
\AFFtit
\AFFtus
\AFFtoyama
\AFFtriumf
\AFFtsinghua
\AFFwu
\AFFwarwick
\AFFwinnipeg
\AFFynu

\author{Y.~Jiang} 
\AFFuci
\author{K.~Abe}
\AFFicrr
\AFFipmu
\author{Y.~Asaoka}
\AFFicrr
\AFFipmu
\author{M.~Harada}
\AFFicrr
\author{Y.~Hayato}
\AFFicrr
\AFFipmu
\author{K.~Hiraide}
\AFFicrr
\AFFipmu
\author{T.~H.~Hung}
\AFFicrr
\author{K.~Ieki}
\author{M.~Ikeda}
\AFFicrr
\AFFipmu
\author{J.~Kameda}
\AFFicrr
\AFFipmu
% ...


\collaboration{The Super-Kamiokande Collaboration}
\noaffiliation

\date{\today}

\begin{abstract}
The detection of solar neutrino provides direct evidence for on-going fusion in the core of the Sun and serve as a probe for that solar core. Production rate of $^8$B is sensitive to the solar core temperature. Adiabatic flavor conversion of $^8$B solar neutrinos depends on the local solar electron density. Combined with the roughly exponential decrease in solar electron density with increasing distance from the center of the solar sphere, the solar core size, density, and temperature are inferred. This article presents an outer bound on solar core size and measurements of the density and temperature in the solar core based on Super-Kamiokande phase I-IV solar neutrino data. 
An outer bound is set for the solar core at $90\%$ confidence level as $0.499R_\odot$ for the high metallicity \smhigh Standard Solar Model and $0.381R_\odot$ for \mzvar which includes protosolar accretion. A combined fit with the SNO polynomial coefficients improves the limit to $0.268R_\odot$ for \smhigh and $0.215R_\odot$ for \mzvar. Results for other metallicity cases are also reported. In the high metallicity scenario assuming nominal solar core density and solar core size, temperature at the solar center is reported at $T_\text{C}=(15.599\pm0.0016)\times10^6$ K and $T_{\text{npz}}=(15.492\pm0.0017)\times10^6$ K at the peak radius of the $^8$B production profile.

\end{abstract}

\maketitle

\textit{Introduction}---Solar neutrinos provide a unique opportunity to make measurements of the Sun's internal structure. Boron eight ($^8$B) solar neutrino are used to probe the size and temperature of the solar core. This letter reports on the upper bound on the solar core and the temperature at the center of the solar sphere and of the neutrino-producing zone measured using the Super-Kamiokande (\SK, SK) phase I-IV recoil electron spectra.\par

The Sun is powered by fusion; the \pp is the primary fusion process. The majority of solar fusion occurs in the center-most region of the Sun, the solar core. Hydrogen burning through \pp fusion accounts for ~99\% of the Sun's power \cite{xsec}. The \pp consists of the three-step process $p+p{\rightarrow}^2\text{H}+e^++\nu_e$, $^2\text{H}+p{\rightarrow}^3\text{He}+\gamma$, and $^3\text{He}+^3\text{He}{\rightarrow}^4\text{He}+2p$ \cite{xsec}, fusing 4 hydrogen nuclei into 1 helium. In addition, the fusion of $^3\text{He}+^4\text{He}{\rightarrow}^7\text{Be}+\gamma$ ($S_{34}$). Production of $^7$Be enables the process $^7\text{Be}+p{\rightarrow}^8\text{B}+\gamma$ ($S_{17}$) . Super-Kamiokande is sensitive to solar neutrinos produced by the $\beta^+$ decay of $^8$B. Neutrino flavor conversion via the Sun's matter effect allows probing of the location of the fusion processes. 

The Mikheyev-Smirnov-Wolfenstein (MSW) effect describes the adiabatic flavor conversion of neutrinos traveling through matter \cite{Mikheyev,wolfenstein}. Full adiabatic conversion to the mass eigenstate $\nu_2$ occurs when neutrinos are produced at a local density exceeding an energy dependent minimum density \cite{Mikheyev}. Solar electron density falls exponentially moving from the solar center outward limiting full adiabatic conversion to the solar core. The effective mixing of mass eigenstates changes with local electron density at the production site and neutrino energy. The survival probability of $^8$B solar neutrinos is therefore sensitive to the density and as a consequence the radial position of production \cite{Zaidel_2025}. This results in a difference in the total number of observed neutrino events for different solar core sizes and density profiles. The energy dependence further enhances the sensitivity through distortion of the recoil electron spectrum.\par

Super-K is a water Cherenkov detector located 1,000 m underground in Kamioka, Japan \cite{Super-Kamiokande:2002weg}. The detector has a total volume of $50$ kton and a fiducial volume of $22.5$ kton. The detector tank consists of two cylindrical layers, the inner detector (ID) and the outer detector (OD). The ID uses 11,129 inward facing 20 inch photomultiplier tubes (PMTs) to detect neutrino events. The layer of water in the OD passively shields against radioactive background. The outward facing PMTs in the OD tag incoming muons; tracked muons are used to reduce spallation background \cite{Abe_2024}.  
Incoming solar $\nu_e$ can elastically scatter with electrons in the water molecules via charged-current (CC) or neutral-current (NC). Incoming $\nu_\mu$ and $\nu_\tau$ can only undergo NC interactions with the electron.If the scattered electron has an energy greater than the Cherenkov threshold, Cherenkov light is emitted. Detected photons are used to reconstruct the energy and the direction of the recoil electron.
 \par


$^8$B fusion and $^8$B neutrino production depend on both $S_{34}$ and $S_{17}$, making the $^8$B neutrino flux sensitive to changes in temperature. Bahcall and Ulmer provide a $T\propto{\Phi_{^8\text{B}}}^{24\pm5}$ power-law relation between the solar center temperature and the total $^8$B flux for evolutionary solar models making $^8$B neutrinos a compelling candidate for precision probing of solar core temperature compared to the $-1.1$ power-law for the \pp neutrinos \cite{powerlaw}. The same power-law relation can be extended to the neutrino production zone defined by the peak of the $^8$B neutrino production profile \cite{Zaidel_2025}.


\textit{Analysis}---This analysis uses \SK recoil electron energy spectrum data \cite{Abe_2024} collected in the full SK I-IV period. Table \ref{tab:skphase} gives a summary of the energy region and livetime of each \SK phase included in this analysis.
Data is reported in the energy range of 3.49 MeV to 19.49 MeV in 23 discrete energy bins, including twenty $0.5$-MeV bins between 3.49--13.49 MeV, two $1$-MeV bins between 13.49--15.49 MeV, and one $4$-MeV bin between 15.49--19.49 MeV. The number of selected events divided by the total number of expected events assuming no oscillation (DATA/SSM) is reported for each bin.\par
\begin{table}
\caption{\label{tab:skphase}Livetime and the energy region for recoil electron energy spectra data during each \SK phase up to SK-IV \cite{Abe_2024}. SK-II has reduced number of PMTs; number of PMTs restored in SK-III; SK-IV the longest period of \SK data-taking under the same detector conditions \cite{Abe:2013gga} and it uses an upgraded electronics system \cite{Super-Kamiokande:2010kjr}.}
\begin{ruledtabular}
\begin{tabular}{lcr}
\textrm{Phase}&
\textrm{Energy region}&
\textrm{Livetime}\\
&\textrm{[MeV]}&\textrm{[days]}\\
\colrule
    SK-I & 4.49-19.49 & 1496\\
    SK-II & 6.49-19.49 & 791\\
    SK-III &3.99-19.49 & 548\\
    SK-IV & 3.49-19.49 & 2970\\
\end{tabular}
\end{ruledtabular}
\end{table}

Choice of solar models: This analysis uses two different solar models, B16 Standard Solar Model \cite{B16} and K2 solar model \cite{k2main} which includes the protosolar accretion phase and a 12\% opacity increase. Two versions of each solar model are analyzed: 
\smhigh is the high metallicity version of the B16 Standard Solar Model, now favored by the Borexino CNO result \cite{cno}, \smlow is the low metallicity version; \mzvar is the K2 model with variable metallicity during accretion, this version most closely reproduces the experimental neutrino fluxes \cite{k2sub}, \czvar with constant metallicity during accretion is also included.\par
The aforementioned temperature analysis is only performed on the two B16 solar models because the power-law relation $T\propto{\Phi_{^8\text{B}}}^{24\pm5}$ was generated based on standard solar models.\par


Solar parameter scaling: The solar core is defined as the region with $90\%$ of all neutrino production in this analysis. The base solar core radius $R_0/R_\odot$ is calculated for each solar model by summing the neutrino production of all species such that $\frac{1}{\Phi_{\text{total}}}\sum_{i}\int^{R_0}_0\Phi_{i}(r/R_{\odot})dr=0.9$ where $\Phi_{i}$ is the neutrino flux of each species \textit{i} in the spherical shell at radius \textit{r}. \par
\begin{figure}
    \includegraphics[width=0.475\linewidth]{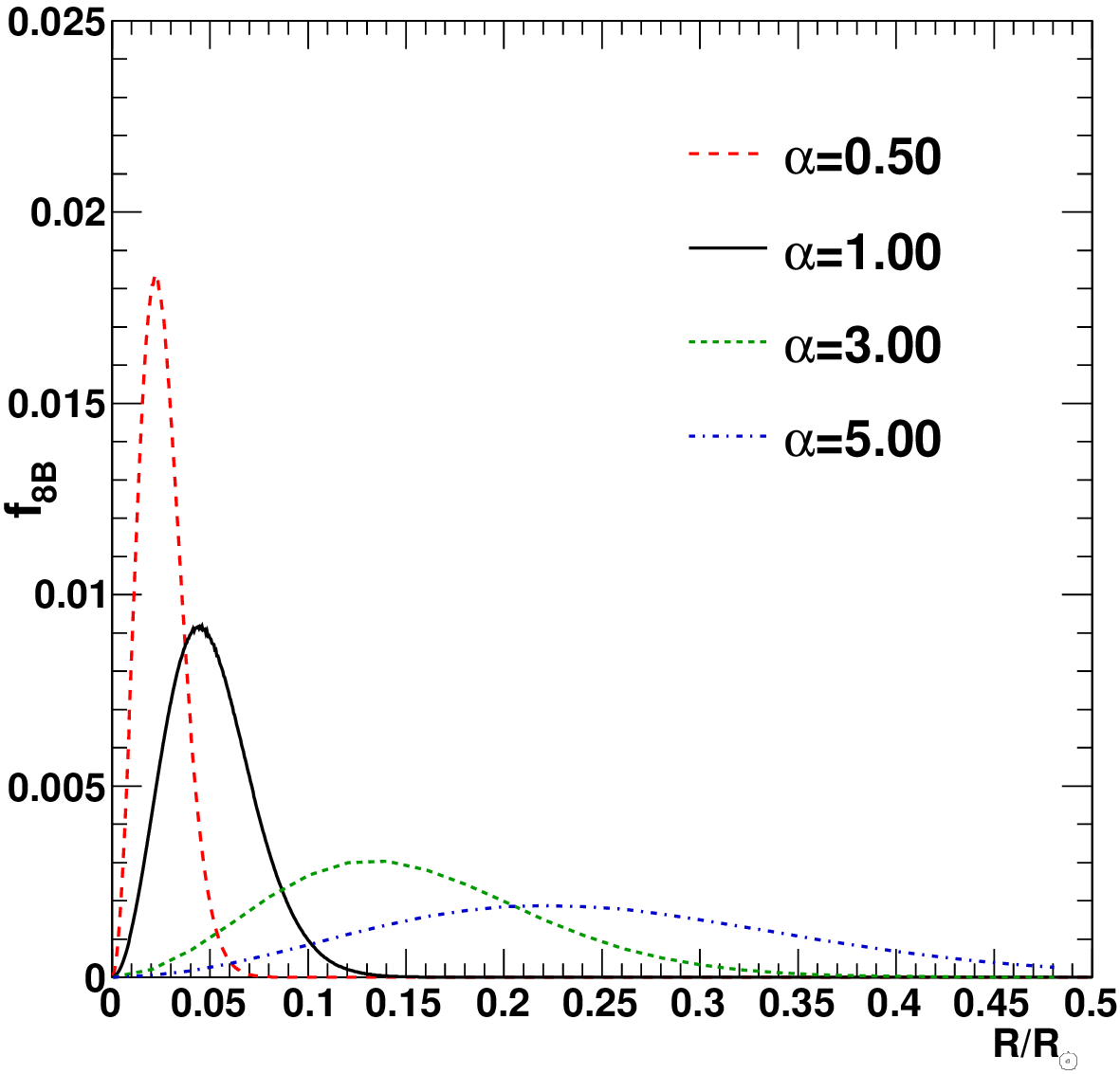}
    \quad
    \includegraphics[width=0.475\linewidth]{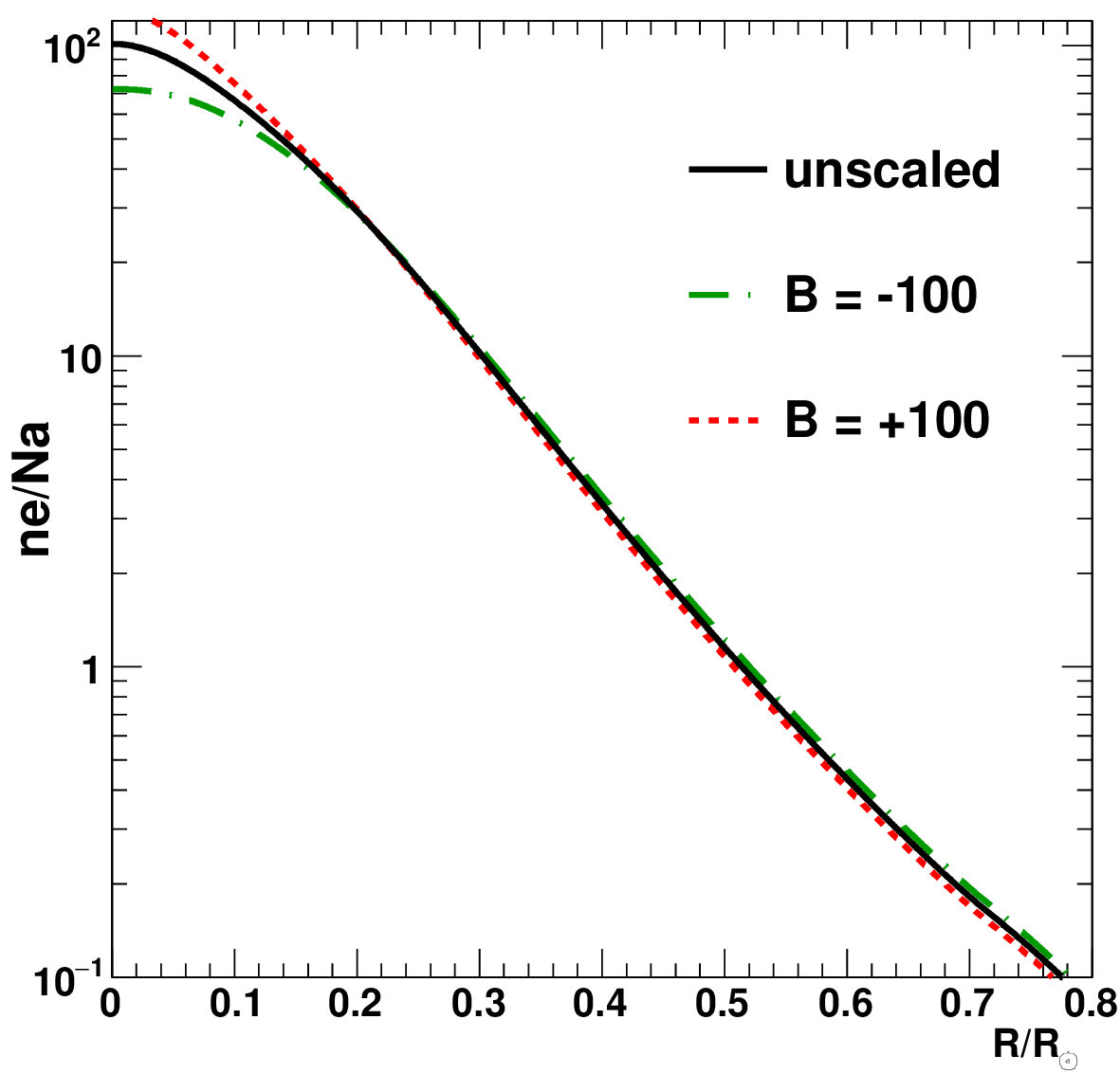}
    \caption{Solar core size and density scaling on \smhigh SSM, using scaling factor $\alpha$ (radius) and \textit{B} (density). Left (a): scaled $^8$B fractional production profile for different $\alpha$ for different solar core size $R_0'=\alpha R_0$. Right (b): unscaled and scaled solar electron number$/$Avogadro's number profile at \textit{B}$=-100,0,100$. }
    \label{fig:scaling}
\end{figure}

Neutrino production as a function of radius $f_{^8\text{B}}$ is normalized: $\int_0^{R_\odot}f_{^8\text{B}}(r/R_\odot)dr=1$
. To test different core sizes, the $^8$B neutrino production profile is scaled by the factor $\alpha$ to form an adjusted core with radius $R_0'$ using Eq.~(\ref{eq:rscale}) such that $R_0'=\alpha R_0$. The modified neutrino production profile $f'_{^8\text{B}}$ retains the same normalization of $f_{^8\text{B}}$. 
\begin{eqnarray}
    f'_{^8\text{B}}=\frac{1}{\alpha}f_{^8\text{B}}(\alpha r/R_\odot)\label{eq:rscale}
\end{eqnarray}
Table~\ref{tab:basecoresize} lists the base solar core radius for each reference solar model used in this analysis. Fig.~\ref{fig:scaling}(a) shows the shapes of the scaled $^8$B profiles at different $\alpha$.\par

To probe the density of the solar core, a second scaling factor \textit{B} is introduced. The electron density profile $\rho(r)$ is deformed by its own shape to form the modified density profile $\rho'(r)$ following Eqs.~(\ref{eq:startden})--(\ref{eq:endden}). 
\begin{eqnarray}
    E&=\displaystyle{\int_0^{R_{\odot}}\rho(r)r^2dr}\label{eq:startden}\\
    I_{\rho}&=\displaystyle{\int_0^{R_{\odot}}\rho^2(r)r^2dr}\\
    \rho'(r)&=[B(\rho(r)/E-I_\rho/E^2)+1]\rho(r)\label{eq:endden}
\end{eqnarray}
This procedure preserves the total electron number $E$ in the Sun while shifting the density relative to the solar center. Positive \textit{B} shifts the density  profile toward the center of the solar sphere
. Fig.~\ref{fig:scaling}(b) shows the electron density profile with no scaling and the profile at both extremes considered in this analysis.\par

\begin{table}
    \caption{\label{tab:basecoresize}Base solar core radius $R_0/R_\odot$ for each reference solar models calculated based on $90\%$ neutrino production of all solar neutrino species.}
   \begin{ruledtabular}
    \begin{tabular}{lc}
    \textrm{Solar Model}     &\textrm{core radius($R_0/R_\odot$)}   \\
    \colrule
    \smlow    & 0.1835\\
    \smhigh   & 0.1830\\
    \mzvar    & 0.1710\\
    \czvar    & 0.1691\\
    \end{tabular}
   \end{ruledtabular}
\end{table}
\SK recoil electron spectrum prediction: Prediction of the \SK recoil electron spectrum is performed using the Super-K Monte Carlo (MC) simulation. The MC is based on skdetsim, a GEANT3-based
 detector simulation with custom photon tracking for Cherenkov photons to the PMTs \cite{Abe_2016}.\par

The $\nu_e$ survival probability is used as input with the \SK MC to predict the recoil electron spectrum at \SK. Table \ref{tab:osc} gives the range of oscillation parameters used for the spectral prediction. The ranges for $\theta_{12}$ and $\Delta m^2_{21}$ are selected to limit computation time. Values allowed by JUNO to at least $4\sigma$ \cite{juno} are considered in this analysis. 
Solar neutrinos have a weak sensitivity to $\theta_{13}$; a fixed value is chosen averaging the central value of Daya Bay \cite{dayabay}, KamLAND \cite{KamLAND}, and JUNO \cite{juno}.\par

\begin{table}
    \caption{\label{tab:osc}Oscillation parameter range used for \SK recoil electron energy spectrum prediction.}
    \begin{ruledtabular}
    \begin{tabular}{ll}
         sin$^2\theta_{12}$  &  [0.166, 0.387]\\
         $\Delta m^2_{21}$   &  [$6.998\times10^{-5}$, $8.091\times10^{-5}$]\\
         sin$^2\theta_{13}$  &  0.02\\
    \end{tabular}
    \end{ruledtabular}

\end{table}

Day-night asymmetry: Neutrinos arriving from below \SK experience matter effects traveling through the Earth. These effects varies with the distance the neutrinos traveled through the Earth leading to a zenith-angle-dependent difference between the day and night neutrino rates described by the day/night asymmetry ($A_{DN}$) \cite{Super-Kamiokande:2013mie}.
\par
The \SK recoil electron spectrum data-MC $\chi^2_{\text{spec}}$ is adjusted based on the $^8$B and $hep$ flux ratios to account for the effect of day-night asymmetry. The recoil electron spectrum is fitted for $\beta$ ($^8$B portion) and $\eta$ (\textit{hep} portion). 
These are combined with the day and night $^8$B and \textit{hep} rates to form the integrated day-night asymmetry for each \SK phases. The combined day-night asymmetry $\chi^2$ of all phases is added to $\chi^2_{spec}$ to form $\chi^2_{spec+ADN}$.\par


\sno polynomial coefficient fit: 
The neutral current, charged-current, and elastic scattering results of \sno I-III are reported in Ref.~\cite{sno123} with six-parameters: three for the oscillation parameters, two for day-night asymmetry, and one for total $^8$B flux. The SNO parameters carries neutrino energy information with analysis uncertainty and cuts already applied. A polynomial of the six SNO parameters is shifted to account for the difference between the solar model used in this analysis and the BS2005 \cite{Bahcall_2005} solar model used in Ref.~\cite{sno123} and \sno polynomial coefficient prediction is formed for each $\alpha$, \textit{B} combination. $\chi^2_{\text{SNO,poly}}$ is formed by a fit of the predicted coefficient curve to the \sno data curve. A simultaneous minimization of $\chi^2_{\text{SK,spectral}}$ and $\chi^2_{\text{SNO,poly}}$ is performed to form the combined SK+SNO fit.\par

Reactor $\chi^2$ fit: JUNO reactor antineutrino results \cite{juno} is used in this analysis to constrain the values of $\Delta m^2_{21}$ and $\theta_{12}$ assuming CPT invariance. For each $\theta_{12}$ and $\Delta m^2_{21}$, $\chi^2_{spec+ADN}$ is combined with $\chi^2_{\text{reactor}}$ form the JUNO result \cite{juno}. The oscillation parameters are marginalized to form the $\chi^2$ for each solar core radius $R_{\text{core}}/R_{\odot}$.\par
\SK total $^8$B flux: The $\nu_e$ cross-section is roughly six times that of $\nu_\mu$ and $\nu_\tau$ for \SK solar neutrino event rate due to the additional NC component. \sno measures a pure neutral-current event rate by counting neutrons from the neutral-current scattering of $\nu_i+D{\rightarrow}p+n+\nu_i$ \cite{sno123}. \SK total $^8$B flux depends on the value of the assumed oscillation parameters. The \sno neutral-current measurement is the primary contribution to the total $^8$B flux. The \SK recoil electron measurement further reduces the uncertainty of the total $^8$B flux, even when taking variation in oscillation parameters into account. The SK total $^8$B flux is fitted in a range consistent with JUNO uncertainty \cite{juno} for each $\theta_{12}$, $\Delta m^2$ pair in units of the \sno neutral current flux $\Phi_{\text{SNO}nc}$.

\begin{figure}
    \includegraphics[width=1\linewidth]{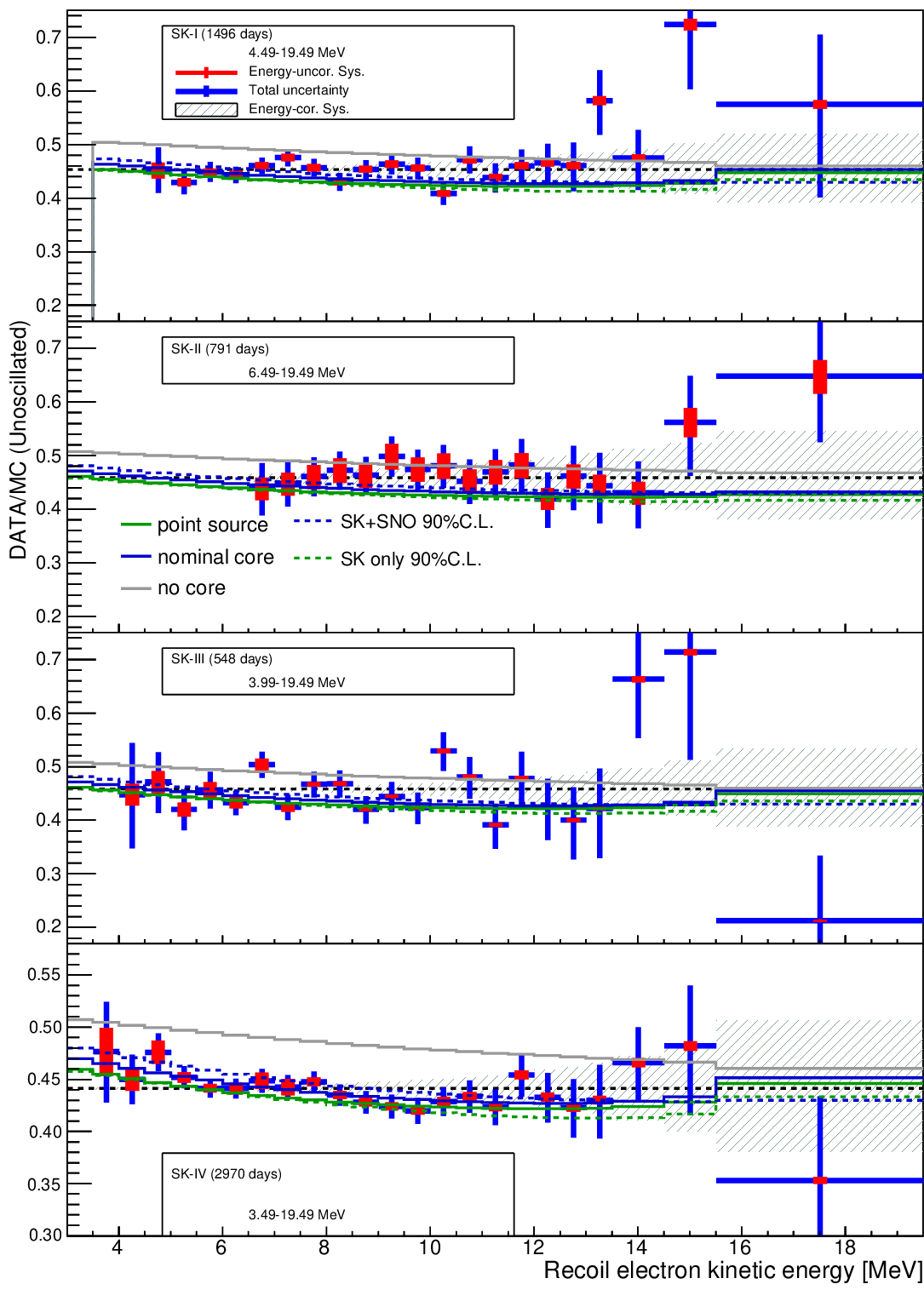}
    \caption{Super-K recoil electron spectrum data and spectral predictions with various core radius scaling $\alpha$ and with density scaling \textit{B}=0. Spectral distortion is applied by allowing variation in $^8$B and hep flux. The no-core case assuming $\nu_\odot$ production zone extending to the surface of the solar sphere is shown by the solid gray line. Energy correlated (fill box around best fit $\Phi_{^8\text{B}}$ ratio) and uncorrelated uncertainty (red bars) are shown separately along with the total uncertainty (blue).}
    \label{fig:specwdata}
\end{figure}

Systematic uncertainties: %
This analysis retains the systematic uncertainties from the standard \SK solar neutrino analysis \cite{Abe_2024}. The uncertainty in this analysis is dominated by the \SK energy scale. Fig.~\ref{fig:specwdata} shows the energy correlated and uncorrelated errors separately. No additional systematic uncertainties have been considered for the day-night asymmetry which is statistics dominated. The SK+SNO fit uses the SNO parameter from Ref.~\cite{sno123} with pre-applied systematic uncertainties.\par

$T_\text{C}$ and $T_{\text{npz}}$: The baseline central temperature $T_{C0}$ and neutrino-production zone temperature $T_{\text{npz}0}$ are obtained from \smlow and \smhigh \cite{B16} solar parameter tables. $T_{\text{C}0}$ is taken at the smallest $r/R_{\odot}$ radial bin. $T_{\text{npz}0}$ is taken at $r/R_{\odot}$ where $f_{^8\text{B}}(r/R_\odot)$ is maximized. The baseline total $^8$B flux $\Phi_{0}$ uses the best-fit $\Phi_{^8\text{B}}$ with no solar core size adjustment ($\alpha=1$).\par

Equation~(\ref{eq:tfrac}) is derived from the power-law relation $T\propto{\Phi_{^8\text{B}}}^{24\pm5}$ provided in Ref.~\cite{powerlaw} assuming the relation holds for modified neutrino production profiles. The temperature $T_{\text{C}0}$ and $T_{\text{npz}0}$ each solar core size hypothesis are calculated using Eq.~(\ref{eq:tfrac}). 
$n$ is varied in accordance to the theoretical limit in the range 19--29 \cite{powerlaw}.\par

\begin{eqnarray}
    \frac{T}{T_0}=\frac{\Phi_{^8\text{B}}^n}{\Phi_0^n}\label{eq:tfrac}
\end{eqnarray}

It should be noted that the fluctuations of temperature in the core due to internal motion, such as solar g-mode oscillations, are expected to be small \cite{Hatta:2026scb}. Hence, those effects are ignored in this analysis.

\begin{figure}
    \includegraphics[width=1\linewidth]{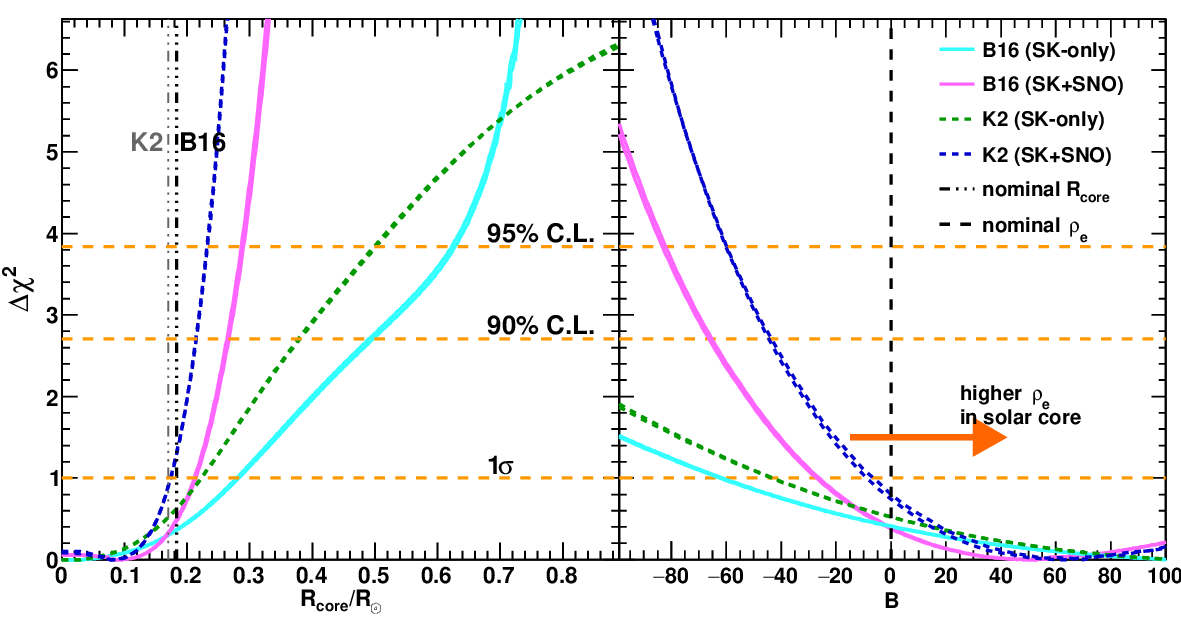}
    \caption{Left (a): $\Delta\chi^2$ distribution as a function of core size $R_{\text{core}}/R_\odot$ with density scaling \textit{B}=0. Right (b): $\Delta\chi^2$ distribution as a function of \textit{B} in the range [$-$100,100] assuming nominal core size $\alpha=1$.}
    \label{fig:x21d}
\end{figure}

\textit{Results}---The upper bound for the solar core radius is determined from the $\Delta\chi^2$. 
Figure~\ref{fig:specwdata} shows the \SK recoil electron spectrum data and a selection of spectral predictions to which it is fitted. 
Figure~\ref{fig:x21d}(a) the 1D $\Delta\chi^2(R_{\text{core}}/R_\odot)$ assuming nominal density. 
The SK+SNO fit shows a slight preference of a non-point-source ($R_{\text{core}}>0$) for all considered solar models. The best fit $R_{\text{core}}$ are $0.092R_\odot$ (\smhigh/\smlow), $0.080R_\odot$ (\mzvar), and $0.082R_\odot$ (\czvar). 
Table~\ref{tab:Rresults} lists the upper bounds to solar core size for each solar model. Difference in metallicity constrains shows negligible effect on the final results.\par

\begin{table}
    \caption{\label{tab:Rresults}Upper bound for solar core size at 90\% C.L. and at 1$\sigma$ for each solar model as fraction of the total solar radius $R_\odot$.}
    \begin{ruledtabular}
      \begin{tabular}{lrr|rr}
      \hspace{1.8cm}$R_{\text{core}}/R_{\odot}$&SK data&&SK+SNO&\\
      &90\%C.L.&1$\sigma$&90\%C.L.&1$\sigma$\\
      \colrule
      \hspace{0.2cm}\smhigh & 0.499& 0.284& 0.268& 0.215\\
      \hspace{0.2cm}\smlow  & 0.494& 0.275& 0.265& 0.212\\
      \hspace{0.2cm}\mzvar  & 0.381& 0.224& 0.215& 0.174\\
      \hspace{0.2cm}\czvar  & 0.374& 0.223& 0.215& 0.174\\

    \end{tabular}
   \end{ruledtabular}
\end{table}

\begin{figure}
    \includegraphics[width=1\linewidth]{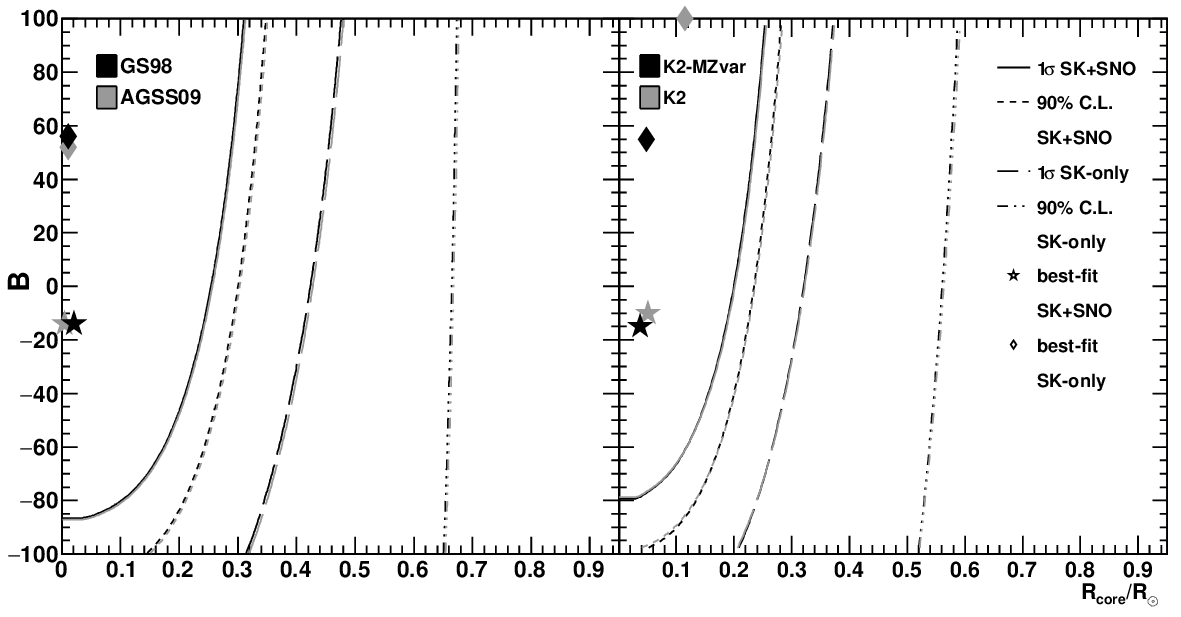}
    \caption{2D $\Delta\chi^2$ contours as a function of core size $R_{\text{core}}/R_\odot$ and density scaling \textit{B}. Positive \textit{B} shifts density toward the center of the solar sphere. Left (a): \smhigh and \smlow. Right (b): \mzvar and \czvar. 
    }
    \label{fig:fig:x22d}
\end{figure}

Figure~\ref{fig:x21d}(b) illustrates the 1D $\Delta\chi^2(B)$ assuming nominal solar core size. The result is consistent with nominal density for all four models with a slight preference for higher density in the core. An lower bound for the density scaling factor \textit{B} is determined using the SK+SNO fit at 90\% confidence level (C.L.) at $B=-67$ for B16 and $B=-45$ for K2.
 Figure~\ref{fig:fig:x22d} shows the contours for the 2D $\Delta\chi^2$ distributions with core size and density scaling. A weak preference for non-point-source at nominal or higher core density is shown. The contours are consistent across all considered models taken into account the difference in nominal solar core size used.\par

\begin{figure}
    \includegraphics[width=1\linewidth]{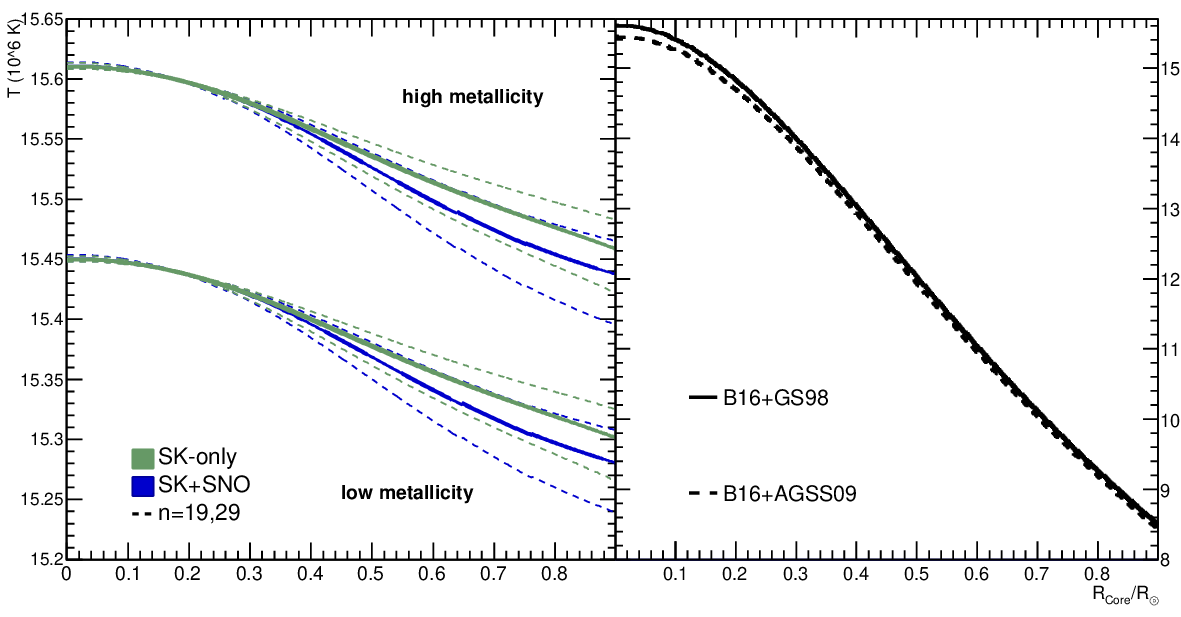}
    \caption{Left (a): solar center temperature $T_\text{C}$ where the thickness of the band shows the uncertainty and the dashed lines shows the theoretical boundary given in Ref.~\cite{Bahcall_2005}. Right (b): neutrino production zone temperature $T_{\text{npz}}$ as a function of core size $R_{\text{core}}/R_\odot$.}
    \label{fig:temp}
\end{figure}

Figure~\ref{fig:temp} shows $T_\text{C}(R)$ and $T_{\text{npz}}(R)$ assuming $n=24$ in the power-law relation. Assuming nominal solar core size, in the SK-only fit for \smhigh, $T_\text{C}=(15.599\pm0.0016)\times10^6$ K and $T_{\text{npz}}=(14.492\pm0.0017)\times10^6$ K. Table~\ref{tab:tresults} lists the temperature results with nominal solar core size ($\alpha=1.00$) and with the SK+SNO best fit solar core size ($\alpha=0.50$).\par

$T_\text{C}$ decreases with $R_{\text{core}}/R_\odot$ due to the reduction in total $^8$B neutrino flux. For $T_{\text{npz}}$, the reduction in local temperature due to a larger neutrino-production zone radius becomes the dominant effect.\par

\begin{table}
    \caption{\label{tab:tresults}Temperature results at the solar center ($T_\text{C}$) and of the neutrino-production zone defined at the peak of the $^8$B neutrino production profile ($T_{\text{npz}}$).}
   \begin{ruledtabular}
    \begin{tabular}{lrrr}
       &$T_\text{C}$ ($10^6$K)&$T_{\text{npz}}$ ($10^6$K)&$R_{\text{npz}}/R_\odot$  \\
    \colrule
    $\alpha=1.00$&&&0.046\\
    SK data only &&&\\
    \hspace{0.15cm}\smhigh   & 15.599$\substack{+0.0015\\-0.0016}$&14.942$\substack{+0.0017\\-0.0017}$&\\
    \hspace{0.15cm}\smlow  & 15.441$\substack{+0.0017\\-0.0018}$&14.798$\substack{+0.0016\\-0.0017}$&\\
    SK+SNO&&&\\
    \hspace{0.15cm}\smhigh   & 15.599$\substack{+0.0012\\-0.0008}$&14.942$\substack{+0.0011\\-0.0007}$&\\
    \hspace{0.15cm}\smlow  & 15.439$\substack{+0.0012\\-0.0007}$&14.798$\substack{+0.0012\\-0.0007}$&\\
    \hline
    $\alpha=0.50$&&&0.023\\
    SK data only &&&\\
    \hspace{0.15cm}\smhigh   & 15.608$\substack{+0.0015\\-0.0016}$&15.436$\substack{+0.0015\\-0.0015}$&\\
    \hspace{0.15cm}\smlow  & 15.447$\substack{+0.0015\\-0.0015}$&15.279$\substack{+0.0015\\-0.0015}$&\\
   SK+SNO&&&\\
    \hspace{0.15cm}\smhigh   & 15.608$\substack{+.0007\\-0.0009}$&15.436$\substack{+0.0007\\-0.0009}$&\\
    \hspace{0.15cm}\smlow  & 15.448$\substack{+0.0006\\-0.0009}$&15.280$\substack{+0.0007\\-0.0009}$&\\
    \end{tabular}
  \end{ruledtabular}
\end{table}

\textit{Conclusion}---The solar core size, density, and temperature are probed using the \SK recoil electron energy spectrum. At 90\% C.L., the analysis based on \smhigh Standard Solar Model gives an outer bound for the solar core at $0.499R_\odot$ in the SK-only fit $0.268R_\odot$ in the SK+SNO fit; analysis based on \mzvar solar model gives an outer bound at $0.381R_\odot$ in the SK-only fit and $0.215R_\odot$ in the SK+SNO fit. The core density is consistent with solar models in all cases considered in this analysis. The solar core temperature is probed based on the B16 SSM using the total $^8$B neutrino flux. Assuming nominal solar core size, in the high metallicity case, the temperature at the solar center to be $T_\text{C}=(15.599\pm0.0016)\times10^6$ K and the temperature of the neutrino production zone to be $T_{npz}=(14.492\pm0.0017)\times10^6$ K. 


\textit{Acknowledgements}---
We gratefully acknowledge the cooperation of the Kamioka Mining and Smelting Company. The SuperKamiokande experiment has been built and operated from funding by the Japanese Ministry of Education, Culture, Sports, Science and Technology; the U.S. Department of Energy; and the U.S. National Science Foundation. Some of us have been supported by funds from the National Research Foundation of Korea (NRF-2009-0083526, NRF-2022R1A5A1030700, NRF-2022R1A3B1078756, RS-2025-00514948, RS-2025-23525600, RS-2026-25489835) funded by the Ministry of Science, Information and Communication Technology (ICT); the Institute for Basic Science (IBS-R016-Y2); and the Ministry of Education (2018R1D1A1B07049158, 2021R1I1A1A01042256, RS-2024-00442775); the Japan Society for the Promotion of Science; the National Natural Science Foundation of China (Grants No. 12375100 and 12521007); the Spanish Ministry of Science, Universities and Innovation (grant PID2024-157541NB-I00); the Natural Sciences and Engineering Research Council (NSERC) of Canada; the Scinet and Digital Research of Alliance Canada; the National Science Centre (UMO2022/46/E/ST2/00336) and the Ministry of Science and Higher Education (2023/WK/04), Poland; the Science and Technology Facilities Council (STFC) and Grid for Particle Physics (GridPP), UK; the European Union's Horizon Europe Research and Innovation Programme HORIZON-MSCA-SE-2023JENNIFER3 project, grant agreement no.101183137, H2020-MSCA-RISE-2019 SK2HK grant agreement no.872549; and European Union's Next Generation EU/PRTR grant CA3/RSUE2021-00559; the National Institute for Nuclear Physics (INFN), Italy. We thank Masanobu Kunitomo from Kurume University for providing the electron density profiles of his latest solar models. This work is partially supported by MEXT KAKENHI Grant Numbers 24K00654

\nocite{*}
\bibliography{ref}

\end{document}